\documentclass[twocolumn,showpacs,amsmath,amssymb]{revtex4}
 
\usepackage{graphicx}
\usepackage{dcolumn}
\usepackage{bm}

\begin{document}
\newcommand{\be}{\begin{equation}}
\newcommand{\ee}{\end{equation}}
\newcommand{\bea}{\begin{eqnarray}}
\newcommand{\n}{\nonumber\\}
\newcommand{\eea}{\end{eqnarray}}

\title{\bf Vulnerability and Protection of Critical Infrastructures}

\author{ Vito Latora$^{1}$ and Massimo Marchiori$^{2,3}$ }

\affiliation{$^{1}$ Dipartimento di Fisica e Astronomia, 
                    Universit\`a di Catania, and INFN, 
                Via S. Sofia 64, 95123 Catania, Italy}
\affiliation{$^{2}$  W3C and Lab. for Computer Science,
               Massachusetts Institute of Technology, 
               545 Technology Square, Cambridge, MA 02139, USA}

\affiliation{$^{3}$  Dipartimento di Informatica, Universit\`a di Venezia, 
Italy }

\date{\today}

\begin{abstract}
Critical infrastructure networks are a key ingredient of modern society.   
We discuss a general method to spot the critical components of a  
critical infrastructure network, i.e. the nodes and the links 
fundamental to the perfect functioning of the network. 
Such nodes, and not the most 
connected ones, are the targets to protect from terrorist attacks.  
The method, used as an improvement analysis, can also help to 
better shape a planned expansion of the network. 
\end{abstract}
{\it Corresponding author: Vito Latora, latora@ct.infn.it}
\vspace{0.5cm}
\maketitle
The attacks of September 11 2001 have raised 
in all its urgency the problem of protecting   
critical infrastructures from terrorist attacks. 
The US President's Commission on Critical Infrastructures Protection 
\cite{pccip} 
has defined five different categories of critical infrastructures: 
1) information-communication, 
2) banking and finance, 
3) energy (e.g. electric, oil, gas), 
4) physical distribution (including transportation), 
5) vital human services (including water supply). 
In this paper we propose a general method to find the 
{\it critical components} of a critical infrastructure 
network \cite{bologna}. 
By critical components we mean the nodes and the edges 
crucial to the best functioning of the network, and 
therefore the strategic points of the network to improve or to 
protect from terrorist attacks. 
Recently, attacks on artificially generated random and scale-free topologies 
and on real-world networks have been studied intensively. In the 
literature appeared so far the attacks were simulated as the deliberate 
removal either of nodes \cite{barabasi3att,holme,clmr1,albert,clm3} or  
of links \cite{girvan,holme,motter} of the network. 
The rationale of our method is different from the previous 
ones: instead of sorting and removing the nodes in descending 
order of degree \cite{barabasi3att,holme,clmr1,albert} 
or betweenness \cite{holme,albert,clm3}, and the edges in descending order 
of betweennes \cite{girvan,holme} or range \cite{motter}, 
we measure the importance of an element of the network 
by the drop in the network's performance caused by the 
deactivation of that element. In practice we check for the 
redundancy of an element by calculating the performance 
of the perturbed network and comparing it with the 
original one. 
The element can be a single node or edge, 
or a group of nodes and edges if we want to simulate multiple attacks. 
In this way we define the vulnerability of 
the network under a given class of attacks and we produce a list of 
the points of the network that should be the first concern of any 
policy of protection from terrorist attacks. 
Analogously, we measure the importance of an improvement by 
the increase in the network's performance caused by such improvement.  
\\
The paper is organized as follows: we first 
present the general framework to define 
critical damages, critical improvements, structural vulnerability 
and improvability of a critical infrastructure. 
We then show how the method works in practice on some examples  
of communication and transportation critical infrastructures. 

\bigskip
We assume that a generic critical infrastructure 
$S$ is characterized by a single 
variable $\Phi[S]>0$, the {\it performance} of $S$ \cite{ext}.
The definition and quantitative analysis of the {\it critical components} 
of $S$, we propose in this paper, uses, as 
reference observable, variations in the performance $\Delta \Phi$.  
We consider separately the study of damages and of  
improvements.

{\bf Attacks analysis}. 
Let us indicate by $D$ a set of possible damages on the 
infrastructure $S$, and with $DAMAGE(S,d)$ a map that gives 
the infrastructure resulting from $S$ after the damage 
$d \in D$. 
We measure the importance of the damage $d$ by the relative drop 
in performance $\Delta \Phi^{-} / \Phi$, with 
$\Delta \Phi^{-} = \Phi[S] - \Phi[DAMAGE(S,d)] \ge 0$,  
caused by $d$. 
In particular, the {\it critical damage} $d^* \in D$ is the damage of $D$ 
that minimizes $\Phi[DAMAGE(S,d)]$. 
The {\it vulnerability} $V$ of $S$ under the 
class of damages $D$ can be defined as: 
\bea
\label{vulnerability}
V[S,D] = \frac{ \Phi[S] - W[S,D]} { \Phi[S] }
\eea
where $W[S,D] = \Phi[DAMAGE(S,d^*)]$ is the worst performance 
of $S$ under the class of damages $D$.  
The vulnerability $V[S,D]$ is defined in the range [0,1].

{\bf Improvements analysis.}
We now turn our attention into how to improve an existing 
infrastructure \cite{twoimprovements}. 
Various improvements can be added to $S$, so given a  
set of improvements $I$ we define, for any improvement $i \in I$,  
the map $IMPROVE(S,i)$ that gives the resulting infrastructure 
obtained after the improvement $i$. 
We measure the importance of $i$ as the relative 
increase in the performance $\Delta \Phi^{+} / \Phi$, 
with $\Delta \Phi^{+}=  \Phi[IMPROVE(S,i)] - \Phi[S]$,  
caused by $i$. 
In particular we define the {\it critical improvement} $i^*$ 
as the best possible improvement in $I$, i.e. 
the improvement of $I$ that maximizes 
$\Phi[IMPROVE(S,i)]$. 
Then, the {\it improvability} $IM$ of $S$ under the 
class of improvements $I$ can be defined as: 
\bea 
\label{improvability1}
IM[S,I] = \frac{ B[S,I]  -  \Phi[S] } { \Phi[S] }
\eea
where $B[S,I]= \Phi[IMPROVE(S,i^*)]$ is the best performance of $S$ under 
the class of improvements $I$.

\bigskip
As a practical application of the method we consider 
communication-information (as the Internet \cite{rosato}) 
and transportation infrastructure networks. We represent the    
infrastructure network $S$ as a valued \cite{wasserman}  
undirected \cite{directed} graph 
with $N$ nodes (for instance the routers in the Internet, or the 
stations in a railway transportation system) and $K$ links (the 
cables connecting two routers, or the lines connecting couples 
of stations). 
$S$ is described by the $N \times N$ adjacency matrix 
$\{ l_{ij} \}$. If there is a link between node $i$ and node $j$, 
the entry $l_{ij}$ is a positive number measuring the {\it link latency}, 
otherwise $l_{ij} =+\infty$. 
For instance, in the Internet (in the railway system) 
the larger  $l_{ij}$ is, the longer it takes for a unitary packet of 
information (a train) to go along the link from $i$ to $j$.  
We have now different ways to measure the performance of $S$. 
In this paper we identify the performance of $S$ with the 
{\em efficiency\/} of the network i.e. we assume: 
$\Phi[S] = E[S] \equiv  \frac{1}{N(N-1)}
{\sum_{{i \ne j\in S}} \frac{1}{d_{ij}}}$, 
where $d_{ij}$ is the smallest sum of the links latency throughout all the 
possible paths in the graph from a node $i$ to a node $j$ 
(in the particular case of unvalued graphs  
$d_{ij}$ reduces to the minimum number of links 
traversed to get from $i$ to $j$). 
The {\it efficiency} is a quantity recently introduced in refs.\cite{lm24} to  
measure how efficiently the nodes of the network communicate if they 
exchange information in parallel. 
A second possibility is to assume the performance $\Phi[S]$ 
to be equal to the inverse of the characteristic path length 
$L \equiv  \frac{1}{N(N-1)}   {\sum_{{i \ne j\in S}} {d_{ij}}}$ \cite{watts,lm24}. 
An alternative possibility to avoid the shortest path assumption 
on which both $E$ and $L$ rely, is to identify $\Phi[S]$ with the mean 
flow-rate of information over $S$ \cite{flow}.

%
\begin{figure}[!h]
\includegraphics[width=8.cm]{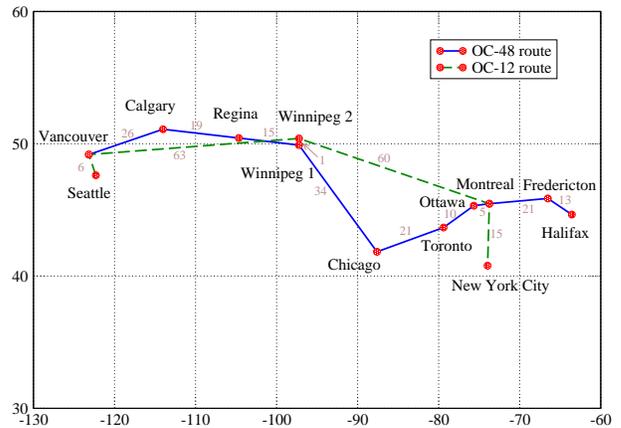}   
\caption{\label{fig1} 
Ca*net3 IS-IS routing network. The numbers reported are 
a measure of the latency associated to each link \cite{canet3}.} 
\end{figure}

{\bf Ca*net3}
We show how the method works in practice by considering 
the Ca*net3 IS-IS routing network \cite{canet3} 
represented in fig.1, a simple example of an Internet 
backbone, consisting of two main routes, OC-12 and  OC-48, 
$N=13$ routers and $K=14$ links.  
As the backbone has diverse routes of different 
bandwidths, the preferred path between any two routers is the 
path which presents the least amount of latency under normal 
router load conditions. 
We consider three different classes (sets) of damages $D$: 
the damage of a single cable connection, 
of a single Internet router, and of a couple of routers. 
$DAMAGE(S,d)$ is the network we obtain 
from $S$ after the deactivation of the damaged component 
(respectively the damaged link, node or couple of nodes). 
The damage of single links allows to investigate the finer  
effects on the network, since the damage of a node 
implies the damage of a number of links equal to the node's  
degree.  
The entity of the damage $d$ is given by the relative drop in 
the efficiency $\Delta \Phi^{-} / \Phi[S]$ caused by $d$.  
\\ 
As class of improvements $I$ we consider the effect of adding 
a new link (the addition of groups of links will be considered in 
\cite{nextpaper}). $IMPROVE(S,i)$ is the network we obtain from 
$S$ after the addition of the new link.
The results shown in table \ref{table1} indicate that the connection 
Winnipeg2-Winnipeg1 is by far the most important one since it is  
crucial for the correct interplay of the OC-12 and OC-48 routes. 
%
\begin{table}
\caption{Attacks and improvement analysis of  Ca*net3. 
For each class of damage/improvement considered (see text) 
we report the cases having the highest effects on the 
performance of the network. 
Rank and name of the damaged link (node, or couple of nodes, respectively) 
and of the added link are listed in the first two columns. 
The relative drop or increase of the 
efficiency is in the third column. 
\label{table1}} 
\begin{tabular}{l|l|l}
& {\em Damaged link\/} & $\Delta \Phi^{-}/\Phi$ \\
  1&  Winnipeg2 -        Winnipeg1      &      0.358  \\
  2&  Ottawa    -        Montreal       &      0.146  \\
  3&  Montreal  -        Fredericton    &      0.123  \\
  4&  Seattle   -        Vancouver      &      0.098  \\
\tableline
&  {\em Damaged node\/} & $\Delta \Phi^{-} / \Phi$  \\
    1 &  Winnipeg1         & 0.466 \\
    2 &  Winnipeg2         & 0.408 \\
    3 &  Montreal          & 0.317 \\
    4 &  Ottawa            & 0.220 \\
\tableline
&  {\em Damaged couple of nodes\/} & $\Delta \Phi^{-} / \Phi$\\
    1  &  Winnipeg1  +  Montreal     &     0.792 \\
    2  &  Winnipeg1  +  Ottawa       &     0.723 \\
    3  &  Winnipeg2  +  Montreal     &     0.702 \\
    4  &  Winnipeg2  +  Ottawa       &     0.700 \\
    5  &  Winnipeg2  +  Toronto      &     0.633 \\
\tableline
\tableline
\tableline
& {\em Added Link\/} & $\Delta \Phi^{+} / \Phi$           \\
   1 & Toronto     -      NYC            &      0.01237\\ 
   2 & Ottawa      -      NYC            &      0.00770\\ 
   3 & Winnipeg1   -      Toronto        &      0.00587\\ 
   4 & Fredericton -      NYC            &      0.00546\\ 
   5 & Winnipeg2   -      Toronto        &      0.00514\\ 
   6 & Seattle     -      Calgary        &      0.00508\\ 
\end{tabular}
\end{table}
The routers Winnipeg1 and Winnipeg2 are respectively the first and 
the second in the list of the most important nodes. 
Conversely when two nodes are removed at once, the couple 
Winninipeg1 + Montreal produces a larger effect than the 
couple Winnipeg1 + Winnipeg2 which is only the tenth in the list 
(not in table) with $\Delta \Phi^{-} / \Phi=0.570$. 
Concerning the improvement analysis, the best links to add 
are long cables bridging two different parts of the network, as 
for instance Toronto-NYC  or Winnipeg1-Toronto.

{\bf Infonet} 
As a second example we study the Internet backbone of 
Infonet\cite{infonet}, as of September 2001.  
The network of Infonet has $N=94$ nodes and $K=96$ cable connections and 
carries about the $10 \%$ of the traffic over US and Europe. 
It consists of two main parts, the US and the European backbone 
respectively with $N_1=66$ and $N_2=28$ nodes, connected by three 
overseas cables. 
In table \ref{table2} we consider the same classes of damages and 
improvements as in the previous example. 
%
\begin{table}
\caption{Attacks and improvement analysis of Infonet 2001 \cite{infonet}, as of 
September 2001. Same as in table \ref{table1}.   
In the last column we report the 
betweenness $b$ of the removed edge, the degree $k$ of the removed node, 
and the sums of the degrees of the two removed nodes. 
\label{table2}} 
\begin{tabular}{l|l|l|l}
& {\em Damaged link\/} & $\Delta \Phi^{-}/\Phi$ & $b$ \\
1&  NYC-New Jersey          & 0.379   & 2205\\      
2&  New Jersey-Chicago      & 0.229   & 1185\\ 
3&  NYC-Washington          & 0.197   & 1185\\ 
4&  Washington-Atlanta      & 0.183   & 1120 \\            
5&  New Jersey-San Jose     & 0.179   & 984\\ 
6&  New Jersey-Dallas       & 0.122   & 609\\
\tableline
&  {\em Damaged node\/} & $\Delta \Phi^{-} / \Phi$ & $k~~~~~$ \\
1 &  New Jersey~~~~~~~~~~    & 0.573 & 9  \\
2 &  NYC ~~~~~~~~~~          & 0.530 & 9  \\
3 &  Chicago ~~~~~~~~~~      & 0.280 & 15 \\
4 &  Amsterdam ~~~~~~~~~~    & 0.241 & 9  \\
5 &  Atlanta   ~~~~~~~~~~    & 0.227 & 14 \\
6 &  Washington ~~~~~~~~~~   & 0.203 & 2  \\
\tableline
&  {\em Damaged couple of nodes\/} & $\Delta \Phi^{-} / \Phi$ & $k_1 +k_2~~~~~$ \\
1 &  NYC~~~~~~~~     +     New Jersey &       0.723 &   17              \\
2 &  New Jersey      +     Amsterdam  &       0.710 &   18              \\
3 &  New Jersey      +     Atlanta    &       0.707 &   23              \\
4 &  New Jersey      +     Frankfurt  &       0.689 &   20              \\
5 &  NYC~~~~~~~~     +     Chicago    &       0.685 &   24              \\
6 &  New Jersey      +     Washington &       0.673 &   11              \\
\tableline
\tableline
\tableline
& {\em Added Link\/} & $\Delta \Phi^{+} / \Phi$ &    \\
1&  New Jersey-Atlanta         &   0.0522       &   \\ 
2&  Chicago-Atlanta            &   0.0481       &   \\
3&  NYC-Atlanta                &   0.0437       &   \\
4&  San Jose-Atlanta           &   0.0395       &   \\
5&  Dallas-Atlanta             &   0.0341       &   \\
6&  Chicago-Amsterdam          &   0.0339       &   \\
7&  NJersey-Amsterdam          &   0.0329       &   \\
8&  NYC-Chicago                &   0.0326       &   \\
9&  Atlanta-Amsterdam          &   0.0318       &   \\
10&  Chicago-Frankfurt         &   0.0316       &   \\
11&  Atlanta-Frankfurt         &   0.0296       &   \\ 
\end{tabular}
\end{table}
The vulnerability of Infonet under single link damages   
is $V=0.379$, with NYC-New Jersey being the critical link damage. 
Such a link plays in the network a role similar to red bonds in 
percolation \cite{percolation}. 
In fact the removal of such a link will result 
in a break up of the network into two disconnected parts of about the same size, 
with a decrease of the $38\%$ in the performance of the network. 
Notice that the second highest link damage produces only a drop 
of $23\%$ in the performance.  
Other important links are those connecting New-Jersey with Chicago, 
with San Jose and with Dallas, and some links in the east cost as  
NYC-Washington and Washington-Atlanta.  
The links in table, ordered according to $\Delta \Phi^{-}/ \Phi$,   
have also a decreasing betweenness $b$, another measure of link centrality 
\cite{wasserman} defined as the number of times the 
link is in the shortest paths connecting couples of nodes \cite{girvan}. 
Nevertheless, the correlation between $\Delta \Phi/^{-} \Phi$ and $b$ 
is not perfect: for instance the link NYC-Amsterdam, with 
the second highest betweenness, 
ranks only 14th according to $\Delta \Phi^{-}/ \Phi$. 
The vulnerability under damages of single nodes (couples of nodes) 
is $V=0.573$ ($V=0.723$). 
New Jersey and NYC are by far the two most important nodes: 
the damage of either one would disconnect the US from the European backbone, 
reducing by more than $50\%$ the performance of the network. 
The damage of both nodes at once reduces by more than $70\%$ the network 
performance. 
The damage analysis of other networks \cite{maps} 
shows that the link NYC-New Jersey and the nodes NYC and 
New Jersey play an important role also in other Internet backbone 
maps. 
Such result might explain the significant drop in performance, 
marked by increased packet loss and difficult in reaching some 
Web Sites (in particular in the connection from 
US to Europe), experienced by the Internet 
in the aftermath of the 11 September terrorist attacks. 
In fact the stress the US Internet infrastructure was subjected to 
was the greatest encountered over its 32-year history  
and was probably related to the damages of Internet 
routers and cables in the south of NYC \cite{media}.
%
\begin{table}
\caption{Attacks and Improvement analysis of the MBTA. 
Same as in table \ref{table1}. The letters in parenthesis indicate 
the line/lines the stations belong to: R=red, G=green, $G_B$=green B, 
$G_C$=green C, O=orange, B=blue.
\label{table3}} 
\begin{tabular}{l|l|l}
   & {\em Damaged link\/} & $\Delta \Phi^{-}/\Phi$ \\
  1&   Park Street(RG)- Boylstone(G)       & 0.275 \\ 
  2&   Boylstone(G)   - Arlington(G)       & 0.270 \\ 
  3&   Arlington(G)   - Copley(G)          & 0.270 \\ 
  4&   Copley(G)      - Hynes(G)           & 0.256 \\ 
  5&   Hynes(G)       - Kenmore(G)         & 0.255 \\ 
  6&   Kenmore(G)     - Blandfor(G)        & 0.185 \\
\tableline
   &  {\em Damaged node\/}           & $\Delta \Phi^{-} / \Phi$ \\
  1&      Kenmore(G)                 & 0.343  \\ 
  2&      Copley(G)                  & 0.333  \\
  3&      Park Street(RG)            & 0.331  \\
  4&      Boylstone(G)               & 0.285  \\
  5&      Arlington(G)               & 0.281  \\
  6&      Hynes(G)                   & 0.266  \\
\tableline
&  {\em Damaged couple of nodes\/} & $\Delta \Phi^{-} / \Phi$ \\
    1&   Down. Cross.(RO)  +  Kenmore(G)      &  0.508 \\
    2&   Park Street(RG)   +  Kenmore(G)      &  0.495 \\
    3&   Down. Cross.(RO)  +  Copley(G)       &  0.465 \\
    4&   Boylstone(G)      +  Kenmore(G)      &  0.444 \\
\tableline
\tableline
& {\em Added Link\/} & $\Delta \Phi^{+} / \Phi$   \\
   1 &  Mount Hood($G_B$)-     Dean ($G_C$)    &  0.0390 \\
   2 &  Mount Hood($G_B$)-     Tappan($G_C$)   &  0.0370 \\
   3 &  Washington($G_B$)-     Tappan($G_C$)   &  0.0369 \\
   4 &  Washington($G_B$)-     Dean ($G_C$)    &  0.0368 \\
   5 &  Sutherland($G_B$)-     Englewood($G_C$)&  0.0360 \\
   6 &  Mount Hood($G_B$)-     Englewood($G_C$)&  0.0357 \\
   7 &  Sutherland($G_B$)-     Dean ($G_C$)    &  0.0355 \\
\end{tabular}
\end{table}
\\
The comparison of our measure with the node 
degree $k$ \cite{wasserman} i.e. with the number of links 
incident with the node, 
(see tab.\ref{table2}) shows that the damage of the most connected 
nodes, the hubs \cite{barabasi3att}, is not always the worst damage. 
In fact, the damage of Chicago, the node with the highest $k$,  
produces only a drop of $28\%$ in the performance 
of the network, while the damage of Chicago and Atlanta,   
the couple with the highest number of links (29) gives  
$\Delta \Phi^{-}/\Phi = 0.476$ (the 187th damage in the list). 
This has deep consequences on the best strategy to 
adopt in a protection policy. 
In fact, a node with a large degree is immediately recognized 
as a major channel of communication,  
being very visible since in direct contact to many 
other nodes \cite{wasserman}. On the other hand, Infonet is a 
typical example in which the crucial components, i.e. the nodes to protect 
from the attacks, are not the hubs, but   
less visible and apparently minor nodes. 
\\
Our results imply either an intense policy of protection of the 
critical links/nodes from attacks, or a strategic expansion of the 
network with the addition of new links \cite{bologna}. 
We now investigate the best strategies to increase the 
performance of the network by the addition of a new link. 
The improvability of $S$ under such a class of 
improvements is $I=0.052$. In the highest positions  
we find two different classes of links:  
links connecting two IP presences in the US, 
and links connecting US and Europe as 
Chicago-Amsterdam, NJersey-Amsterdam, Atlanta-Amsterdam, 
Chicago-Frankfurt and Atlanta-Frankfurt.  
A new link between Us and Europen, namely the link Washington-Geneva,  
was in fact planned in the expansion of Infonet 2001.  
Our method predicts that the inclusion of such a link 
increases by $2.5 \%$ the network performance. 

{\bf MBTA}
As a final example we cosider a transportation system, the 
Boston subway, consisting of four lines, $N=124$ stations and $K=125$ 
tunnels \cite{mbta}. Here the links latency has been taken to be proportional 
to the time it takes to go from a station to the next one.
The results of the analysis are in table \ref{table3}. 
The vulnerability $V$ is equal to 0.275,0.343,0.508, respectively for damages 
of single links, single nodes or couples of nodes. The critical link is 
Park Street - Boylstone. $I$ is equal to 0.0390 with best links to be added 
those connecting stations on the green line B with stations on the 
green line C.

Summing up, in this paper we have proposed a new general method to 
spot the critical components of a critical infrastructure system. 
With this method we are able to identify the points of a network 
that are crucial to the functioning of the infrastructure network,  
i.e. those nodes and connections whose protection from terrorist 
attacks must be assumed as the first concern of any national policy. 
The method, used as an improvement analysis, can also help to 
better shape an expansion of the network. 
Other classes of critical 
infrastructure systems are currently under study and will be presented 
in a future work \cite{nextpaper}.


\small
\begin{thebibliography}{99}

\bibitem{pccip} 
http://www.info-sec.com/pccip/pccip2/index.html

\bibitem{bologna}
S. Bologna, C. Balducelli, G. Dipoppa and G. Vicoli, 
Lecture Notes in Comp. Sci. {\bf 2788}, 342 (2003).

\bibitem{barabasi3att} R. Albert, H. Jeong, and A.-L. Barab\'asi,
{\it Nature\/} {\bf 406}, 378 (2000); 
{\it Nature\/} {\bf 409}, 542 (2001). 

\bibitem{holme}
P. Holme, B. J. Kim, C. N. Yoon and S. K. Han, 
{\it Phys. Rev. E\/} {\bf 65}, 056109 (2002).

\bibitem{clmr1}
Crucitti, P., V. Latora, M. Marchiori, and A. Rapisarda. 
{\it Physica A} {\bf 320}, 622 (2003).

\bibitem{albert} R. Albert, I. Albert and G.L. Nakarado,
{\it Phys. Rev. E\/} {\bf 69}, 025103(R) (2004).

\bibitem{clm3} P. Crucitti, V. Latora and M. Marchiori,
{\it Phys. Rev. E\/} {\bf 69}, 045104(R) (2004). 

\bibitem{girvan} 
M. Girvan, and M. E. J. Newman. 
{\it Proc. Natl. Acad. Sci. USA} {\bf 99}, 8271 (2002).

\bibitem{motter} 
A. E. Motter, T. Nishikawa, Y. Lai,  
{\it Phys. Rev. E\/} {\bf 66}, 065103 (2002). 

\bibitem{ext} 
The method can be extended to the case in which 
the performance is a combination of two 
or more variables.  

\bibitem{twoimprovements}
There are two main improvement strategies: we can 
better shape the expansion of a given infrastructure 
in order to increase its performance, or in order to decrease 
its vulnerability. The most general strategy is an 
appropriate combination of the above two strategies, to 
get a good mixture of performance and low vulnerability. 
In this paper we adopt the first of the two strategies. 

\bibitem{rosato}
V. Rosato, F. Tiriticco, 
{\it Europhys. Lett.\/} {\bf 66}, 471 (2004). 

\bibitem{wasserman} S. Wasserman and K. Faust,
{\it Social Networks Analysis} (Cambridge University Press,
Cambridge, 1994).

\bibitem{directed} The formalism presented in this paper can be 
easily extended also to {\it directed} graphs. 


\bibitem{lm24} V. Latora and M. Marchiori,
{\it Phys. Rev. Lett.} {\bf 87}, 198701 (2001); 
{\it Eur. Phys. J.} {\bf B32}, 249-263 (2003).

\bibitem{watts} D.J. Watts and S.H. Strogatz,
{\it Nature} {\bf 393}, 440 (1998).

\bibitem{flow} L.R. Ford and D.R. Fulkerson, 
{\it Flows in networks} (Princeton University Press,
Princeton, 1962).

\bibitem{canet3} 
http://205.189.33.72/optical/pdf/canet3routing.pdf

\bibitem{nextpaper} P. Crucitti, V. Latora and M. Marchiori, 
in preparation.  

\bibitem{infonet}
http://www.infonet.com

\bibitem{percolation} 
H.E. Stanley, {\it J. Phys. A} {\bf 10}, 1211 (1977). 

\bibitem{maps}
http://navigators.com/isp.html

\bibitem{media} 
http://www.cnn.com/2001/TECH/industry/09/12/telecom.operational.idg/  

\bibitem{mbta} 
http://www.mbta.com/





\end{thebibliography}
\end{document}